\documentclass[aps,prl,twocolumn,superscriptaddress,floatfix]{revtex4-1}
\usepackage{graphicx}
\usepackage{color}

\usepackage{ifpdf}

\ifpdf
  \usepackage[pdftex]{graphicx}
  \DeclareGraphicsExtensions{.pdf}
  \usepackage[a4paper,dvipdfm,hyperindex=true]{hyperref}
\else
  \usepackage{graphicx}
  \DeclareGraphicsExtensions{.ps,.eps}
  \usepackage[a4paper,dvipdfm,hyperindex=true]{hyperref}
\fi

\definecolor{linkcol}{rgb}{0.2,0.2,0.6}

\hypersetup
{
bookmarksopen=true,
pdfmenubar=true, 
pdfhighlight=/O, 
colorlinks=true, 
pdfpagemode=None, 
pdffitwindow=true, 
linkcolor=linkcol, 
citecolor=linkcol, 
urlcolor=linkcol 
}

\def\TN{\ensuremath{T_{\rm N}}}
\def\Qb{\ensuremath{|\mathbf{Q}|}}
\def\SCTO{Sr$_2$CuTeO$_6$}

\begin{document}

\title{Magnetic Excitations and Electronic Interactions in Sr$_2$CuTeO$_6$: \\ A Spin-1/2 Square Lattice Heisenberg Antiferromagnet }

\author{P. Babkevich}
\email{peter.babkevich@gmail.com}
\affiliation{Laboratory for Quantum Magnetism, Institute of Physics, \'{E}cole Polytechnique F\'{e}d\'{e}rale de Lausanne (EPFL), CH-1015 Lausanne, Switzerland}
\author{Vamshi M. Katukuri}
\email{vamshi.katukuri@epfl.ch}
\affiliation{Chair of Computational Condensed Matter Physics, Institute of Physics, \'{E}cole Polytechnique F\'{e}d\'{e}rale de Lausanne (EPFL), CH-1015 Lausanne, Switzerland}
\author{B. F{\aa}k}
\affiliation{Institut Laue-Langevin, CS 20156, F-38042 Grenoble Cedex 9, France}
\author{S. Rols}
\affiliation{Institut Laue-Langevin, CS 20156, F-38042 Grenoble Cedex 9, France}
\author{T. Fennell}
\affiliation{Laboratory for Neutron Scattering and Imaging, Paul Scherrer Institut, CH-5232 Villigen, Switzerland}
\author{D. Paji\'{c}}
\affiliation{Department of Physics, Faculty of Science, University of Zagreb, Bijeni\v{c}ka c. 32, HR-10000 Zagreb, Croatia}
\author{H. Tanaka}
\affiliation{Department of Physics, Tokyo Institute of Technology, Meguro, Tokyo 152-8551, Japan}
\author{T. Pardini}
\affiliation{Lawrence Livermore National Laboratory, Livermore, California 94550, USA}
\author{R.~R.~P. Singh}
\affiliation{Department of Physics, University of California, Davis, California 95616, USA}
\author{A. Mitrushchenkov}
\affiliation{ Laboratoire Mod\'{e}lisation et Simulation Multi Echelle, MSME UMR 8208 CNRS, Universit\'{e} Paris-Est, 5 bd Descartes, 77454 Marne-la-Vall\'{e}e, France}
\author{O. V. Yazyev}
\affiliation{Chair of Computational Condensed Matter Physics, Institute of Physics, \'{E}cole Polytechnique F\'{e}d\'{e}rale de Lausanne (EPFL), CH-1015 Lausanne, Switzerland}
\author{H. M. R\o nnow}
\affiliation{Laboratory for Quantum Magnetism, Institute of Physics, \'{E}cole Polytechnique F\'{e}d\'{e}rale de Lausanne (EPFL), CH-1015 Lausanne, Switzerland}

\begin{abstract}
\SCTO\ presents an opportunity for exploring low-dimensional magnetism on a square lattice of $S=1/2$ Cu$^{2+}$ ions. We employ \emph{ab initio} multi-reference configuration interaction calculations to unravel the Cu$^{2+}$ electronic structure and to evaluate exchange interactions in \SCTO. The latter results are validated by inelastic neutron scattering using linear spin-wave theory and series-expansion corrections for quantum effects to extract true coupling parameters. Using this methodology, which is quite general, we demonstrate that \SCTO\ is an almost ideal realization of a nearest-neighbor Heisenberg antiferromagnet but with relatively weak coupling of 7.18(5)\,meV.
\end{abstract}

\date{\today}
\maketitle

Mott insulators are a subject of intense interest due to the observation of many different quantum phenomena~\cite{khomskii-book, witczak-krempa-review-2014}. In low-dimensional systems, frustration and quantum fluctuations can destroy long-range magnetic order giving rise to quantum paramagnetic phases such as valence-bond solids with broken lattice symmetry or spin liquids, where symmetry is conserved but with possible new collective behaviors involving emergent gauge fields and fractional excitations
\cite{anderson-1973, anderson-science-1987, balents-nature-2010}.
 The spin-1/2 frustrated square-lattice with nearest-neighbor (NN) $J_1$ and next-nearest neighbor $J_2$ exchange interactions is one of the simplest models for valence-bond solids and spin liquids~\cite{anderson-science-1987,shirane-prl-1987}. Yet, despite the many theoretical efforts, experimental realizations of the $J_1$-$J_2$ model have been rather scarce.
The double perovskite oxides are particularly interesting as magnetic interactions can be tuned by changing structure, stoichiometry and cation order \cite{king-jmc-2010, vasala-prog-2015}. In the search for a quantum magnet with weak exchange energies, \SCTO\ has been proposed \cite{iwanaga-jssc-1999, koga-jpn-2014}.

\begin{figure}[t]

\includegraphics[width=1\columnwidth]{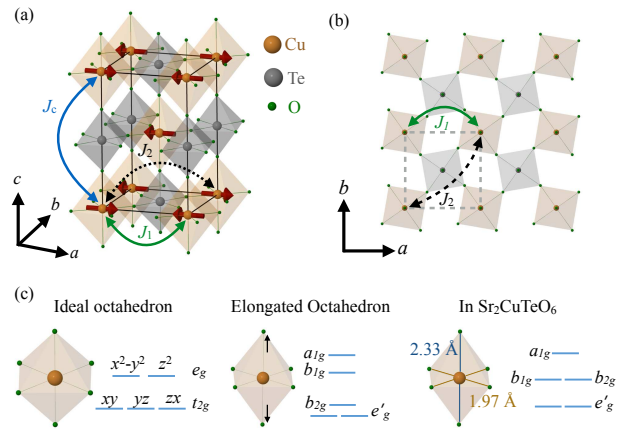}

\caption{(a) and (b) Crystallographic and magnetic structure of \SCTO. The Cu$^{2+}$ ions order magnetically into an arrangement indicated by the red arrows. The different exchange couplings are shown by arrows connecting two Cu$^{2+}$ ions.
(c) Energy level diagram of $d$-states in octahedral ($O_h$) symmetry, for a tetragonally elongated octahedron and for the Cu$^{2+}$ ion in \SCTO\ whose Cu-O bond lengths are labeled. }
\label{fig1}
\end{figure}

The tetragonal crystal structure of the double perovskite \SCTO\,\cite{Reinen_SCTO_Struct_ZAAC}
consists of corner sharing CuO$_6$ and TeO$_6$ octahedra that are rotated in a staggered fashion
about the $c$-axis; see Figs.~\ref{fig1}(a) and \ref{fig1}(b). The CuO$_6$ octahedra are elongated along the $c$-axis,
effectively resulting in the ground state of a Cu$^{2+}$ ($3d^9$) ion having a hole in the in-plane $d_{x^2-y^2}$ orbital, where $z$ is along the $c$-axis.
This could eventually result in quasi-2D magnetism in \SCTO\, with dominant intra-plane exchange interactions.
In the basal $ab$-plane, the exchange that couples the Cu$^{2+}$ ions is the super-superexchange interaction mediated through the bridging TeO$_6$ octahedra as shown in Fig.~\ref{fig1}(b), which is expected to reduce the coupling strength in \SCTO.

Magnetic susceptibility and heat capacity measurements on \SCTO\, indicate a quasi-2D magnetic behavior, suggesting that it is a realization of the square-lattice $J_1$-$J_2$ model~\cite{koga-jpn-2014}. More recently, neutron diffraction measurements on \SCTO\ have shown it to order in a N\'{e}el antiferromagnetic (AFM) structure below $\TN\simeq29$\,K with moments in the $ab$-plane~\cite{koga-prb-2016}; see Fig.~\ref{fig1}(a). The ordered moment at 1.5\,K was found to be reduced to 0.69(6)\,$\mu_{\rm B}$, from the classical value of 1\,$\mu_{\rm B}$~\cite{koga-prb-2016}, indicating a renormalization by quantum fluctuations \cite{reger-prb-1988, singh-prb-1989}. These observations demand further investigation into the magnetic ground state and excitations that elucidate the role of quantum effects in \SCTO.

In this Letter, we show that \SCTO\ is an almost ideal realization of a two-dimensional square lattice Heisenberg antiferromagnet. This is achieved by a novel {\em ab initio} configuration interaction calculation of relevant exchange interactions, which are reaffirmed by modeling the inelastic magnetic spectrum using spin-wave theory and correcting the exchange interactions by series expansion.


Let us first consider the electronic interactions in \SCTO. For a Cu$^{2+}$ ($3d^9$) ion in O$_6$ octahedral ligand cage, the degenerate $3d$ levels are split into low-energy $t_{2g}$ and high-energy $e_g$ manifolds with a hole in the latter. In the tetragonally elongated CuO$_6$ octahedra in \SCTO, the degeneracy of $t_{2g}$ and $e_g$ is further reduced into states with $e^\prime_g$, $b_{2g}$ ($t_{2g}$), and $b_{1g}$, $a_{1g}$ ($e_g$) symmetry as shown in Fig~\ref{fig1}(c).
The ground state wavefunction composition of Cu$^{2+}$ in \SCTO\, and the $d$-level excited state energies and corresponding wavefunctions are summarized in Table~\ref{tab1:dd_exc}. These are obtained from calculations at complete-active-space self-consistent-field (CASSCF) and multireference configuration-interaction (MRCI) levels of the many-body wavefunction theory~\cite{book_QC_00},
\begin{table}[!t]
\caption{
Relative energies of the Cu$^{2+}$ ion $d$-level excitations in \SCTO (in hole representation). The composition of wavefunctions at the CASSCF level is also provided. Only the five $3d$ orbitals of the Cu$^{2+}$ ion were included in the CASSCF active space. At MRCI level, the wavefunction would also contain contributions from the other correlated orbitals (see text).
}
\label{tab1:dd_exc}
\begin{tabular}{ccc}
\hline
\hline\\[-0.40cm]
Symmetry         &Relative E (eV)      &CASSCF  \\
of $d^9$ states       &CASSCF/MRCI            & wavefunction\\
\hline\\[-0.25cm]
$a_{1g}$  &0.00/0.00         &$0.97\,|d_{x^2-y^2} \rangle + 0.24\,|d_{xy} \rangle $\\
$b_{2g}$  &0.778/0.856         &$-0.24\,|d_{x^2-y^2} \rangle + 0.97\,|d_{xy} \rangle $\\
$b_{1g}$  &0.796/0.863         &$1.0\,|d_{z^2} \rangle $\\
$e^{\prime}_g$  &1.013/1.098         &$0.94\,|d_{yz} \rangle - 0.34\,|d_{zx} \rangle $\\
                &1.013/1.098         &$0.34\,|d_{yz} \rangle + 0.94\,|d_{zx} \rangle $\\[0.10cm]
\hline
\hline
\end{tabular}
\end{table}
on embedded clusters of atoms containing a single reference CuO$_6$ octahedron and the surrounding six TeO$_6$ octahedra; see Supplemental Material \footnote{See Supplemental Material, which includes Refs.~\onlinecite{ewald, ANO_Cu_basis, ANO_O_basis, Te_basis, ANO-S_O_basis, Sr_basis, NOCI_J_hozoi03,Li2Cu2O2_Js,Ir214_katukuri_12, Te_ecp_vtz, crystal_14, vasp, jmol, J_ligand_fink94,J_ligand_calzado03,NOCI_J_oosten96, Ir213_katukuri_13,Ir113_bogdanov_12,Ir214_katukuri_12,Ir214_katukuri_14, localization_PM, VASP_PAW, wannier90, schmidt-prb-2011}} for computational details. In contrast to correlated calculations based on density functional theory
in conjunction with dynamical mean field theory (DFT+DMFT), our calculations are parameter free and accurately describe correlations within the cluster of atoms in a systematic manner.
An active space of nine electrons in five $3d$ orbitals of the Cu$^{2+}$ ion was considered at the CASSCF level to capture the correlations among the $3d$ electrons.
In the subsequent correlated calculation, on top of the CASSCF wavefunction all single and double (MR-SDCI) excitations were allowed from the Cu $3s,3p,3d$ and O $2p$ orbitals of the reference CuO$_6$ octahedron into virtual orbital space to account for correlations involving those electrons~\cite{liviu_sci_rep,Hsiao_Yu_PRB_2011}. All calculations were done using the {\sc molpro} quantum chemistry package~\cite{Molpro12}.

From Table~\ref{tab1:dd_exc} it is evident that, at the CASSCF level, the ground state hole orbital predominantly has $d_{x^2-y^2}$ character with a small $d_{xy}$ component. This admixture is due to the staggered rotation of CuO$_6$ and TeO$_6$ octahedra. Note that the wavefunction obtained in the MR-SDCI calculation also contains non-zero weights from those configurations involving single and double excitations into O $2p$ orbitals.
The MR-SDCI calculations predict the lowest crystal field excitation
($a_{1g}$--$b_{1g}$ and $a_{1g}$--$b_{2g}$)
to be nearly degenerate at 0.86 eV,
an accidental degeneracy very specific to \SCTO. The highest $d$-level excitation is at 1.01 eV; see Fig.~\ref{fig1}(c).
It is interesting to note that the on-site $d$-$d$ excitations in \SCTO\, occur at rather
low energies in comparison with 1D or 2D layered cuprates~\cite{Moretti_rixs_2011,liviu_sci_rep,Hsiao_Yu_PRB_2011}.
The presence of highly charged Te$^{6+}$ ions around the CuO$_6$ octahedron effectively decrease the effect of the ligand field on the Cu $d$-orbitals~\cite{Note1}, a phenomenon observed in layered perovskite compound Sr$_2$IrO$_4$~\cite{Bogdanov15}.

Having established the ground state hole orbital character and Cu$^{2+}$ on-site $d$-$d$ excitations in \SCTO\,, we evaluate the exchange interactions shown in Fig.~\ref{fig1}(a).
The exchange couplings were derived from a set of three different MRCI calculations on three different embedded clusters. To estimate $J_1$ a cluster consisting of two active CuO$_{6}$ octahedral units and two bridging TeO$_6$ octahedra was considered, for $J_2$ and $J_c$ only one bridging TeO$_6$ octahedron was included in the active region~\cite{Note1}.

The coupling constants were obtained by mapping the energies of the magnetic configurations of the two unpaired electrons in two Cu$^{2+}$ ions onto that of a two-spin Heisenberg Hamiltonian ${\mathcal H}_{ij}=J_{ij}\mathbf{S}_i\cdot\mathbf{S}_j$.
A CASSCF reference wavefunction with two electrons in the two Cu$^{2+}$ ground state $d_{x^2-y^2}$-type orbitals was first constructed for the singlet and triplet spin multiplicities~\cite{coen_degraf_book}, state averaged.
In the MRCI calculations the electrons in the doubly occupied Cu $3d$ orbitals and the Te $4d$ and O $2p$ orbitals of the bridging TeO$_6$ octahedron were correlated. We adopted a difference dedicated configuration interaction (MR-DDCI) scheme~\cite{DDCI_1_MIRALLES1992555,DDCI_2_MIRALLES199333} recently implemented within {\sc molpro}, where a subset of the MR-SDCI determinant space
\footnote{In MR-SDCI all configurations involving single and double excitations from the active and inactive orbitals to the virtual orbital space are considered in the wavefunction expansion.} that {\em excludes} all the double excitations from the inactive orbitals to the virtuals, is used to construct the many-body wavefunction. This approach has resulted in exchange couplings for several quasi-2D and quasi-1D cuprates that are in excellent
agreement with experimental estimates, e.g. see Ref. \onlinecite{cuo2_j_illas_00,*cuo2_j_illas_04,*cuo2_dm_pradipto_2012,*j_licu2o2_pradipto_2012}.

In Table~\ref{Exchg_inter} the Heisenberg couplings derived at the CASSCF and MR-DDCI level with Davidson corrections for size-consistency errors~\cite{Davidson_MRCI} are listed. We see that all the interactions are AFM.
At the fully correlated  MR-DDCI level of calculation we obtain the in-plane exchange coupling $J_1$ to be the largest at 7.39 meV.
At the CASSCF level where the Anderson type of exchange is accounted for, i.e, related to intersite $d$-$d$ excitations of the $d_{x^2-y^2}^0$--$d_{x^2-y^2}^{2}$ type~\cite{Anderson_1950,anderson1959}, only 30\% of the $J_1$ exchange is obtained. The MR-DDCI treatment, which now includes excitations of the kind $t_{2g}^5e_g^1$ -- $t_{2g}^6e_g^2$, etc., and O $2p$ to Cu 3$d$ charge-transfer virtual states as well, enhances $J_1$.
Our calculations estimate a second neighbor in-plane coupling $J_2 = 0.007 J_1$ and the coupling along the $c$-axis to be practically zero; see Table~\ref{Exchg_inter}.

Although it may perhaps be expected that the dominant superexchange comes from bridging Te $4d$-orbitals -- the path Cu$^{2+}$-O$^{2-}$-Te$^{6+}$-O$^{2-}$-Cu$^{2+}$, we find that the Te outer most occupied $4d$ orbitals are core-like at $\approx 50$ eV below the valence Cu $3d$ and the oxygen $2p$ orbitals, and, hence, a negligible contribution to the magnetic exchange. A MR-DDCI calculation that does not take into account the virtual hopping through the Te $d$ states results in $J_1 = 7.79$\,meV.
Thus we conclude that the dominant superexchange path is Cu$^{2+}$-O$^{2-}$-O$^{2-}$-Cu$^{2+}$ along the two bridging TeO$_6$ octahedra; see ~\cite{Note1}.
Interestingly, we find that the superexchange involving virtual hoppings from the doubly occupied Cu $3d$ orbitals of $t_{2g}$ symmetry and the $d_{z^2}$ of $e_g$ symmetry, ~ 0.86-1.1\,eV lower than the $d_{x^2-y^2}$ orbitals (see Table~\ref{tab1:dd_exc}), contribute almost half to the exchange coupling -- a calculation without the doubly occupied Cu $d$ orbitals in the inactive space result in a $J_1$ of 4.51\,meV.

\begin{table}[!t]
\caption{
Heisenberg exchange couplings derived from {\em ab initio} CASSCF/MR-DDCI data and experimentally for \SCTO. The experimental in-plane couplings were obtained from fits to INS using SWT and corrected by SE; see text. Values are given in meV.}
\label{Exchg_inter}
\begin{tabular}{cccc}
\hline
\hline\\[-0.40cm]
$J$      & CASSCF      &MR-DDCI & Experimental\\
\hline\\[-0.25cm]
$J_1$  & 2.320 &  7.386 & 7.18(5)\\
$J_2$  & 0.006 &  0.051 & 0.21(6)\\
$J_c$  & 0.000 &  0.003 & 0.04\\
\hline
\hline
\end{tabular}
\end{table}

Next, we turn to inelastic powder neutron scattering (INS) measurements to determine experimentally the nature of magnetic interactions. The experiments were performed at Paul Scherrer Institute, using the spectrometer FOCUS (not shown), and Institut Laue-Langevin on the thermal time-of-flight spectrometer IN4~\cite{cicognani-in4c}.

The data were collected on a sealed Al envelope containing 24.1\,g of \SCTO\ powder at temperatures of 2, 60, and 120\,K with
incident neutron energy of 25.2\,meV and Fermi chopper at 250\,Hz. The raw data were corrected for detector efficiency, time-independent background, attenuation, and normalized to a vanadium calibration following standard procedures using LAMP and Mslice software packages \cite{lamp, *mslice}.

\begin{figure}[b]
\includegraphics[bb= 20 30 544 455, clip=,width=1\columnwidth]{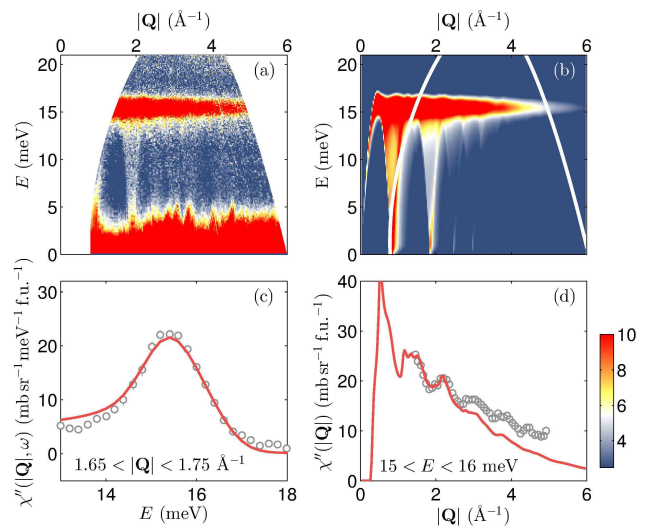}
\caption{
(a) Dynamic susceptibility $\chi''(\Qb,E)$ map obtained by subtraction of 120\,K from 2\,K data. (b) Calculated powder average inelastic spectrum using $\tilde{J}_1 = 7.60(3)$ and $\tilde{J}_2 = 0.60(3)$\,meV. The solid white lines show the detector edges. The SWT intensity is scaled to match the measured pattern in units of mb\,sr$^{-1}$\,meV$^{-1}$f.u.$^{-1}$. (c) A constant wavevector cut for $\Qb\approx1.7$\,\AA$^{-1}$ through the dynamic susceptibility with a solid red line showing the calculated cut from SWT. (d) \Qb-dependence of the magnetic band around 15.4\,meV and a comparison to SWT calculations.
}
\label{fig2}
\end{figure}


The spin-spin correlations between Cu ions can be probed using INS as a function of momentum and energy transfer $(\Qb,\hbar\omega)$, where the former is defined as $\Qb=|h \mathbf{a}^\ast + k \mathbf{b}^\ast + l \mathbf{c}^\ast|$ in terms of reciprocal lattice vectors. The magnetic neutron scattering cross-section is directly related to the imaginary part of the dynamical susceptibility $\chi''(\Qb,\omega)$. At sufficiently high temperatures above \TN, the magnetic excitations are generally heavily damped and uncorrelated \cite{ronnow-prl-2001}.
In the case of \SCTO, some magnetic correlations persist even at 60\,K ($\approx2\TN$), indicative of the low-dimensionality of the system, see Ref.~\cite{Note1}. On warming to 120\,K, the magnetic signal can no longer be observed and we subtract this data from the 2\,K measurements to reveal a purely magnetic contribution to the signal.

In Fig.~\ref{fig2} we present the measured and calculated magnetic spectra. Figure~\ref{fig2}(a) shows the inelastic powder $\chi''(\Qb,\omega)$ spectrum mapped over momentum and energy transfer. We observe dispersive modes originating from the magnetic Bragg peak positions around $\Qb=0.9$ and 1.87\,\AA$^{-1}$, which correspond to $(0.5,0.5,0)$ and $(1.5,0.5,0)$. The dispersion is linear, which is consistent with AFM spin waves and remains gapless within the energy resolution of our measurements of 1.4\,meV at the elastic line (FWHM).

The dominant feature in our spectrum is a strong, flat band around 15.4\,meV, shown in Fig.~\ref{fig2}(c). The intensity decreases with increasing \Qb\ as expected for magnetic scattering; see Fig.~\ref{fig2}(d). For low-dimensional systems, powder averaging produces a van Hove-like maximum at the zone boundary. Therefore, we interpret the flat band as due to the zone boundaries and not to a dispersionless excitation. We observe that the signal at 15.4\,meV has a FWHM of 1.7\,meV, which is significantly larger than the 1.2\,meV instrumental resolution at this energy transfer. This implies that there is dispersion along the zone boundary.

There are two potential sources of zone boundary dispersion. First, a finite $J_2$ leads to dispersion along the zone boundary. This effect can be captured by spin wave theory (SWT). Second, it has been well established that even the purely nearest neighbor ($J_2=0$) spin-$1/2$ Heisenberg antiferromagnet on a square lattice exhibits a quantum effect with two results: (i) at $(\pi,0)$ the sharp spin-wave peak develops a lineshape extending towards higher energies -- a quantum effect that has often been explained in terms of spinon deconfinement~\cite{dallapiazza-nature-2015}; ii) a 6\%-8\% zone boundary dispersion where $E(\pi,0)$ is lower than $E(\pi/2,\pi/2)$. The latter effect cannot be captured by SWT but by several other theoretical approaches -- series expansion (SE) \cite{singh-prb-1995, zheng-prb-2005}, exact diagonalization \cite{chen-prb-1992}, quantum Monte-Carlo (QMC) \cite{syljuasen-jpcm-2000, sandvik-prl-2001}, variational wave-function (VA) \cite{dallapiazza-nature-2015}, etc.  In the presence of an AFM $J_2$ coupling, the quantum dispersion and the $J_2$ dispersion reenforce each other.

For calculating the powder-averaged neutron spectra, the classical (large-$S$) linear spin-wave (SWT) works best, owing to significantly faster computation time. Therefore, our approach is to fit the magnetic spectrum using SWT to extract effective $\tilde{J}_1$ and $\tilde{J}_2$ parameters and then to use SE to correct these values to obtain true $J_1$ and $J_2$ parameters. In doing so, we consider a Heisenberg Hamiltonian,
$
\mathcal{H} = \tilde{J}_1 \sum_{\langle ij\rangle} \mathbf{S}_i\cdot\mathbf{S}_j + \tilde{J}_2 \sum_{\langle ij\rangle} \mathbf{S}_i\cdot\mathbf{S}_j$.
We neglect the very small $c$-axis coupling as obtained in our calculations, see Table~\ref{Exchg_inter}. The magnetic dispersion can be described as $\hbar\omega = Z_c\sqrt{A^2 - B^2}$, where $A=2\tilde{J}_1 + \tilde{J}_2[\cos(2\pi h - 2\pi k) + \cos(2\pi h + 2\pi k) -2]$ and $B = \tilde{J}_1(\cos 2\pi h + \cos 2\pi k)$
\footnote{The first order quantum correction to SWT is to multiply the calculated dispersion by $Z_c=1.18$ \cite{singh-prb-1989}. In presence of $J_2$, $Z_c$ becomes very weakly $\mathbf{Q}$-dependent. However, for $J_2 \ll J_1$ it is a good approximation to use a constant $Z_c=1.18$.}.
To fit the data we calculate the imaginary part of the dynamic susceptibility including an anisotropic Cu$^{2+}$ magnetic form factor \cite{brown-itc-2006,zaliznyak-book}. The resulting spectrum is shown in Fig.~\ref{fig2}(b) which has been calculated using $\tilde{J}_1 = 7.60(3)$ and $\tilde{J}_2 = 0.60(3)$\,meV. Comparing the spectra in Figs.~\ref{fig2}(a) and \ref{fig2}(b), we find good agreement across the entire wavevector and energy transfer range. The SWT simulation is able to reproduce the strong flat mode around 15.4\,meV and spin-waves emerging from the AFM positions.  At larger \Qb, we find that the intensity is predicted to decrease more rapidly than observed; see Fig.~\ref{fig2}(d). This could be an artifact of imperfect subtraction of the phonon spectrum, a small mixing of the $d_{xy}$ orbitals influencing the magnetic form factor or multiple scattering.

\begin{figure}
\includegraphics[width=1.0\columnwidth]{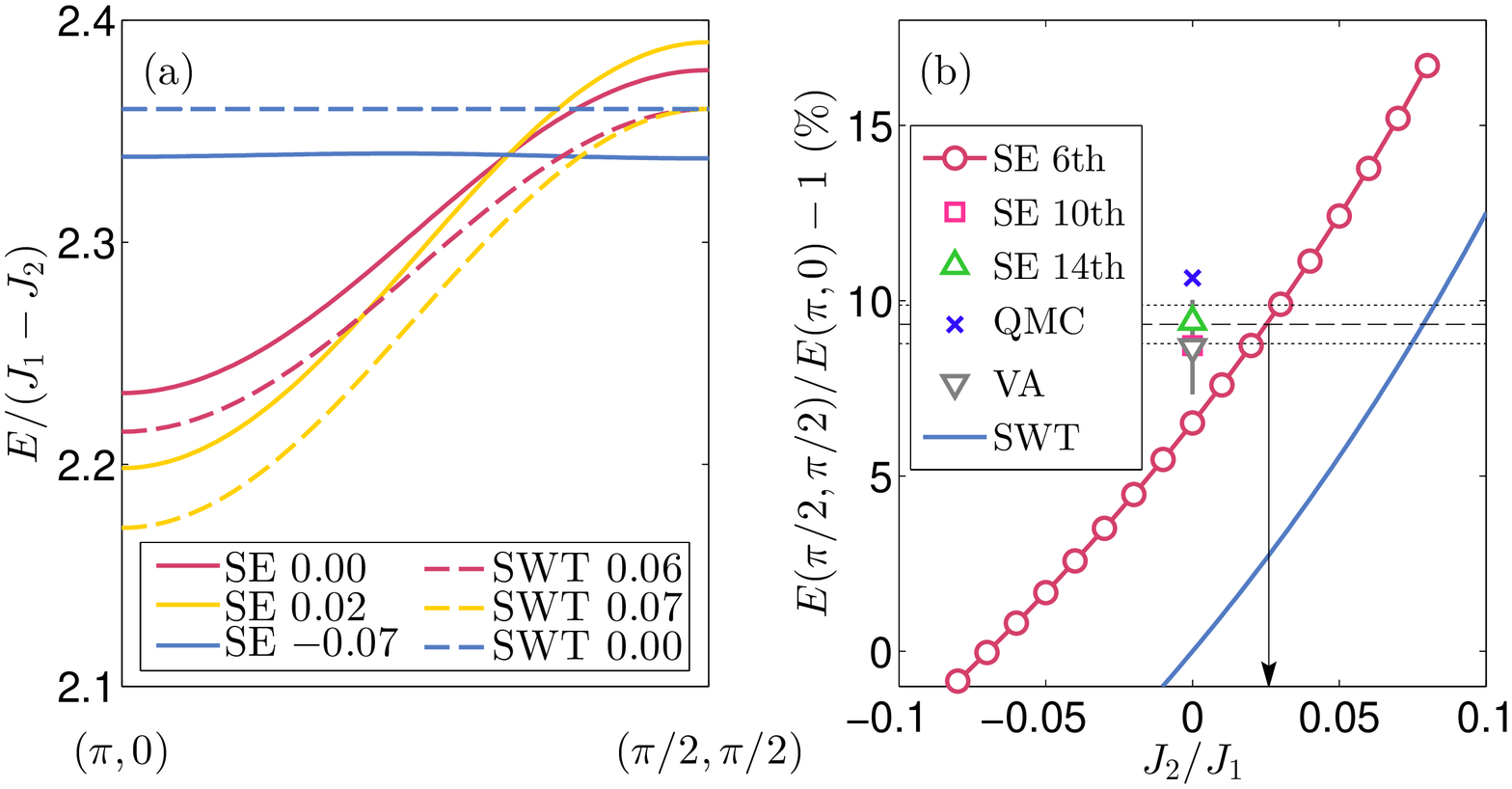}
\caption{
(a) Dispersion between $(\pi,0)$ and $(\pi/2,\pi/2)$ calculated using SE and SWT methods for $J_1=1$ and $J_2/J_1$ values given in the legend. (b) Calculated change of zone boundary energies at $(\pi,0)$ and $(\pi/2,\pi/2)$ obtained from SE and SWT. The dashed horizontal line denotes the dispersion expected for the calculated ratio of $\tilde{J}_2/\tilde{J}_1=0.078$ and the horizontal lines either side show the corresponding uncertainty of $\pm0.005$ in $\tilde{J}_2/\tilde{J}_1$.}
\label{fig4}
\end{figure}

We now turn to the series-expansion method up to 6th order for $J_1$-$J_2$ to correct the exchange coupling parameters derived from SWT for the quantum effects \cite{singh-prb-1995,tsyrulin-prl-2009}. Figure~\ref{fig4}(a) shows the calculated single-magnon energies for the SE and SWT calculations for different relative strengths $J_2/J_1$ and $\tilde{J}_2/\tilde{J}_1$. We employ the convention where $(\pi/2,\pi/2)$ and $(\pi,0)$ correspond to points $(h,k)=(1/2,0)$ and $(1/4,1/4)$ (and equivalent) in reciprocal space, respectively. The SE calculations show a zone-boundary dispersion of around 7\% when second neighbor exchange is absent. Comparing this to SWT calculations, see Fig.~\ref{fig4}(a), it is clear that a non-zero AFM $\tilde{J}_2$ parameter modifies this part of the dispersion in a similar manner.

From SWT fits, we find that $\tilde{J}_2/\tilde{J}_1 = 0.079(3)$ which leads to a 9.3(5)\% dispersion between $(\pi/2,\pi/2)$ and $(\pi,0)$. However, in SE, the same dispersion is explained largely by quantum fluctuations, see Fig.~\ref{fig4}(b), such that $J_2/J_1 = 0.025(5)$, or $J_2=0.21(6)$\,meV. By correcting the SWT results by SE, we obtain a more realistic value of the ratio of the exchange coupling parameters. The zone boundary dispersion can be estimated by other theoretical approaches for $J_2=0$ \cite{dallapiazza-nature-2015,chen-prb-1992,syljuasen-jpcm-2000, sandvik-prl-2001,zheng-prb-2005}. In Fig.~\ref{fig4}(a) we show that the same amount of dispersion as we observe can also be explained in the absence of $J_2$ interaction. Nonetheless, our experimental results place an upper limit on the size of $J_2$. We note that reducing $J_2$ must increase $J_1$ accordingly $J_1 \approx \tilde{J}_1 (1-\tilde{J}_2/\tilde{J}_1)/(1-J_2/J_1)$, which results in $J_1\approx7.18(5)$\,meV.
For a quasi-2D system, \TN\ can be used to estimate the coupling $J_c$ between layers using $\TN\approx J_c [\xi(\TN)/a]^2$. We find the correlation length is $\xi(\TN)/a\approx 10$ from three-loop order given in Ref.~\cite{hasenfratz-plb-1991}. This gives an out-of-plane coupling on the order of 0.04\,meV.
%
Comparing experimentally obtained exchange parameters with {\em ab initio} calculations in Table~\ref{Exchg_inter}, we find remarkably good agreement. Indeed, this demonstrates the power of our approach in obtaining a complete description of the magnetic interactions which has rather rarely been applied to strongly correlated electron systems.

We note that neutron scattering measurements have recently been performed on the related Sr$_2$CuWO$_6$ compound where the $J_2 \gg J_1$ leads to columnar antiferromagnetic order \cite{walker-prb-2016,burrows-arxiv}.
Exchange parameters have been estimated using calculations based on density functional theory corrected for Hubbard type interactions and are in reasonable agreement with experiments without corrections for quantum fluctuations~\cite{walker-prb-2016,burrows-arxiv}.
It would be interesting to validate the proposed exchange interaction mechanisms in Sr$_2$CuWO$_6$ using more accurate many-body calculations similar to those adopted in this work.

In summary, we have characterized magnetic interactions in a new layered antiferromagnet \SCTO\ using detailed {\em ab initio} configuration interaction calculations and inelastic neutron scattering measurements.
The calculations accurately predict the exchange interactions,
and further determine the dominant exchange path i.e via Cu$^{2+}$-O$^{2-}$-O$^{2-}$-Cu$^{2+}$ and not via Te 4$d$ orbitals, as previously suggested.
By simulating the magnetic excitations using classical SWT corrected by SE, we show that NN exchange coupling is around 7.18(5)\,meV with very weak next-nearest interactions on the order of $<3$\% of $J_1$.
The low-energy scale of interactions in \SCTO\ should make it an appealing system to study theoretically and experimentally as an almost ideal realization of a nearest-neighbor Heisenberg antiferromagnet. Moreover, our work brings to the fore a novel strategy for exploring Heisenberg antiferromagnets from {\em ab initio} calculations to simulations of magnetic spectra taking into account quantum effects.

\begin{acknowledgments}
We wish to thank I. \v{Z}ivkovi\'{c}, S. Katrych, L. Hozoi, and N.~A.~Bogdanov for fruitful discussions. P.B. is grateful for help from R.S. Ewings in implementing spherical powder averaging. V.M.K. and O.V.Y. acknowledge the support from ERC project `TopoMat' (Grant No. 306504). This work was funded by the European Research Council grant CONQUEST, the SNSF and its Sinergia network MPBH. Series expansion simulations were performed under the auspices of the U.S. Department of Energy by Lawrence Livermore National Laboratory under Contract No. DE-AC52-07NA27344. Document Release No. LLNL-JRNL-692712. This work was supported by a Grant-in-Aid for Scientific Research (A) (Grant No. 26247058) from Japan Society for the Promotion of Science. D.P. acknowledges partial support of Croatian Science Foundation under the Project 8276.

P. B. and V. M. K. contributed equally to this work.
\end{acknowledgments}

\bibliography{shorttitles,biblio_V8}

\clearpage
\newpage
\appendix

\section{Supplemental Material}

\section{Multi-reference configuration interaction calculations}
\subsection{Ground state of Cu$^{2+}$ ion and $d$-level excitations}

A cluster consisting of a single CuO$_6$ octahedron surrounded with six TeO$_6$ octahedra and the eight Sr ions was considered for calculating the Cu$^{2+}$ ground state and on-site $d-d$ excitations. The surrounding solid-state matrix was modeled as a finite array of point charges fitted to reproduce the crystal Madelung field in the cluster region~\cite{ewald}. We employed all electron atomic natural orbital (ANO) basis sets of quadruple-zeta quality for the central Cu$^{2+}$ ion~\cite{ANO_Cu_basis} and triple-zeta functions for the oxygens~\cite{ANO_O_basis} of the central CuO$_6$ unit. Additionally three polarization $f$-functions for the Cu ion and two $d$-functions for the oxygens were used. The Te ions were represented  by energy-consistent pseudopotentials and triple-zeta basis sets for the valence shells~\cite{Te_basis} and oxygens connected to the Te ions were represented with ANO [$2s1p$] functions~\cite{ANO-S_O_basis}. For the Sr$^{2+}$ ions we used total-ion effective potentials and a single $s$ valence basis function~\cite{Sr_basis}.
\begin{figure}[!hb]
\includegraphics[angle=0,width=1.0\columnwidth]{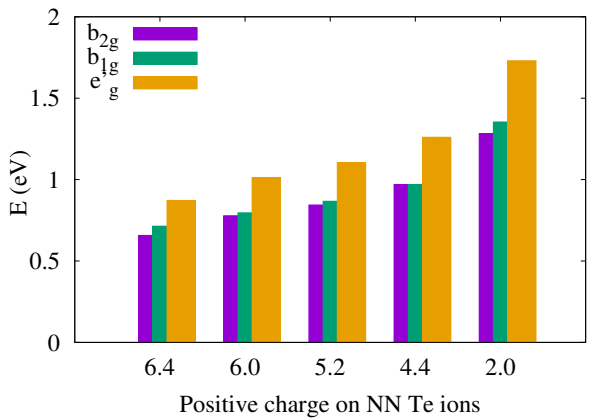}
\caption{(Color online)
Excited state energies of $d$-manifold of the Cu$^{2+}$ ion for varying charges on the NN Te ions.
}
\label{fig_SM_1}
\end{figure}

As mentioned in the main text, the on-site $d$-$d$ excitations in \SCTO\ occur at rather low energies in comparison with other cuprate compounds. To verify that this is due to the highly charged Te$^{6+}$ ions surrounding the CuO$_6$ octahedron, we calculated the $d$-level energies for several scenarios with different charges on the neighboring Te ions. As shown in Fig.~\ref{fig_SM_1}, as the charge on the NN Te ions decreases, the $d-d$ excitation energies increase. For 2+ charge on the NN Te ions, the excitation energies are very similar to other perovskite cuprates, e.g. La$_2$CuO$_4$~\cite{liviu_sci_rep}.

\subsection{Magnetic couplings between two Cu$^{2+}$ ions.}

Three different clusters were adopted to evaluate $J_1$, $J_2$ and $J_c$, see Fig. 1 in the main text. For $J_1$ the cluster consists of two CuO$_6$ and two TeO$_6$ octahedra constituting the reference unit and eight surrounding TeO$_6$ octahedra (buffer region) to describe the charge distribution in the reference region accurately.
The cluster for $J_2$ consists of a reference unit with two CuO$_6$ and one bridging TeO$_6$ octahedra, and the buffer region has ten TeO$_6$ octahedra and four Cu$^{2+}$ ions surrounding the bridging TeO$_6$ octahedron. These Cu$^{2+}$ ions were described by closed shell total ion potentials to avoid spin-couplings with the Cu$^{2+}$ ions in the reference region. Such procedure is often used in quantum chemistry calculations for solids, see Ref.~\onlinecite{NOCI_J_hozoi03,Li2Cu2O2_Js,Ir214_katukuri_12}.
The cluster used for calculating $J_3$ is similar to that of the one used for $J_2$ calculation. All the Sr$^{2+}$ ions surrounding the reference unit are also considered in all the three clusters.
As for the single site calculations, the solid-state matrix was modeled as a finite array of point charges fitted to reproduce the crystal Madelung field in the cluster region.

ANO quadruple-zeta quality basis sets with three polarization $f$-functions were used for the Cu$^{2+}$ ions~\cite{ANO_Cu_basis} of the reference unit and the bridging oxygens were described with quintuple-zeta basis sets and four polarization $d$-functions~\cite{ANO_O_basis}. Triple-zeta quality basis functions~\cite{ANO_O_basis} were used for the rest of the O$^{2-}$ ions in the reference unit. The 28 core electrons of bridging Te ions were represented with effective core potential and the occupied 4$s$, 4$p$, 4$d$ and unoccupied 5$s$, 5$p$ manifolds were represented with [$6s5p4d$] basis functions with two additional polarization $f$-functions~\cite{Te_ecp_vtz}. All the 48 electrons of the Te$^{6+}$ ions in the buffer region were modelled with effective core potential and valence 5$s$ and 5$p$ are described with triple-zeta quality basis sets~\cite{Te_basis}. For the oxygen ions in the buffer region we used ANO type two $s$ and one $p$ function~\cite{ANO-S_O_basis}.

To evaluate the accuracy of our embedding scheme, we have calculated the Ising-like coupling, $J_1^{I}$, for the Cu-Cu link corresponding to $J_1$
using periodic unrestricted Hartree-Fock (UHF) calculations. {\sc crystal}~\cite{crystal_14} program package was employed for these periodic UHF calculations. We used triple zeta basis sets for Cu and oxygen from {\sc crystal} library. Te and Sr
ions were treated with effective core potentials with [2s2p] basis functions for the valence electrons~\cite{Te_basis,Sr_basis}.

Table~\ref{tab-uhf} summarizes the results of such calculations.
The $J_1^{I}$ calculated from periodic calculation should be exactly the same as that obtained from our cluster calculation at UHF level of
calculation and with the same basis sets if the embedding was exact representation of the solid environment.
One can see that there is a difference of 0.8 meV between the periodic and
embedded cluster calculations.

\begin{table} [!b]
\caption{
Ising coupling, $J_1^I$, calculated using unrestricted Hartree-Fock method with perioding and embedded cluster approaches.
}
\label{tab-uhf}
\begin{tabular}{cccc}
\hline
\hline\\[-0.40cm]
UHF      & \multicolumn{2}{c}{ Relative energies (meV) }\\
\cline{2-3}
 & Periodic   & Emb. cluster \\
\hline\\[-0.25cm]
$J_1^I$ &  -0.56 & 0.25 \\
\hline
\hline
\end{tabular}
\end{table}

\begin{figure} 
\includegraphics[angle=0,width=0.8\columnwidth]{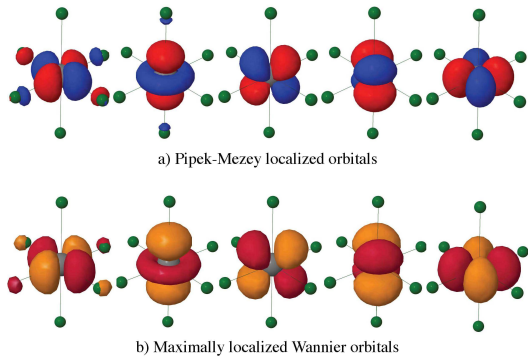}
\caption{(Color online)
Localized Cu $3d$ orbitals, labels from left to right: $d_{x^2-y^2}$, $d_{z^2}$, $d_{xz}$, $d_{yz}$ and $d_{xy}$. a) Orbitals used in our calculations. b) Maximally localized Wannier functions obtained from periodic density functional theory calculations using VASP\cite{vasp} package.
Jmol~\cite{jmol} plotting package was used to plots the orbitals.
}
\label{fig_wfc}
\end{figure}

In the correlated calculations, multiconfiguration reference wave functions were first generated by CASSCF calculations. For two NN CuO$_6$
octahedra, the finite set of Slater determinants was defined in the CASSCF treatment in terms of two electrons and two Cu $d_{x^2-y^2}$ type of orbitals.
The SCF optimization was carried out for an average of the singlet and triplet states associated with this manifold.
On top of the CASSCF reference, the MR-DDCI expansion additionally includes single and double excitations into the virtual orbitals from the reference CAS space and single excitations from the Cu $d$ orbitals not included in the CAS space and the 2$p$ orbitals of the bridging ligands. A similar strategy of explicitly dealing only with selected groups of ligand
orbitals was earlier adopted in quantum chemistry studies on both 3$d$~\cite{J_ligand_fink94,J_ligand_calzado03,NOCI_J_oosten96} and 5$d$~\cite{Ir213_katukuri_13,Ir113_bogdanov_12,Ir214_katukuri_12,Ir214_katukuri_14} compounds, with results in good agreement with the experiment. To separate the Cu 3$d$ and O 2$p$ valence orbitals into different groups, we used the Pipek-Mezey~\cite{localization_PM} orbital localization module available in {\sc molpro}~\cite{Molpro12}.

In Fig. \ref{fig_wfc} the Cu $3d$ orbitals obtained by Pipek-Mezey localization and the Cu 3$d$ maximally localized Wannier functions obtained from a periodic density functional theory calculation are shown. The latter were obtained from calculations performed using VASP~\cite{vasp} package with PAW pseudopotentials~\cite{VASP_PAW} and the Wannier90~\cite{wannier90} code. One can see that the ones used in the cluster calculations are very similar to the maximally localized Wannier functions.

\begin{figure} [!b]
\includegraphics[angle=0,width=0.8\columnwidth]{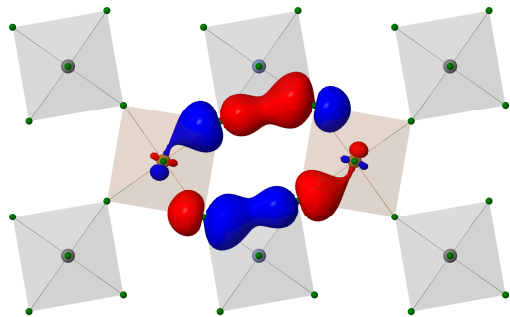}
\caption{(Color online)
The sigma overlapping bridging oxygen 2$p$ orbitals that contribute primarily to the exchange coupling $J_1$. The Cu, O and Te atoms are shown in brown, green and grey colors respectively. Jmol~\cite{jmol} plotting package was used to produce the molecular orbitals.
}
\label{fig_SM_2}
\end{figure}

As mentioned in the main text, we find that the super-superexchange path Cu$^{2+}$-O$^{2-}$-O$^{2-}$-Cu$^{2+}$ along the two bridging TeO$_6$ octahedra contributes to the exchange interaction $J_1$. In Fig.~\ref{fig_SM_2} the bridging O $2p$ orbitals that primarily contribute to the exchange coupling are shown. Table~\ref{table_SM_1} lists the exchange coupling $J_1$ obtained when particular O $2p$ orbitals are correlated. The $p_x$ type of orbitals have a sigma-type overlap with the Cu $d_{x^2-y^2}$ orbitals where as the $p_y$ are nearly orthogonal and the $p_z$ are orthogonal. It can be clearly seen that the sigma overlapping orbitals contribute to the exchange significantly.
\begin{table} [!t]
\caption{
Exchange coupling $J_1$ obtained when selected oxygen bridging $2p$ orbitals are correlated, see text. The values are in meV.
}
\label{table_SM_1}
\begin{tabular}{cccc}
\hline
\hline\\[-0.40cm]
      & \multicolumn{3}{c}{Correlated orbitals}\\
\hline
CASSCF & $p_x$   &$p_y$ & $p_z$ \\
\hline\\[-0.25cm]
2.32 &  6.64 & 3.71& 3.53\\
\hline
\hline
\end{tabular}
\end{table}

\section{Magnetic susceptibility}

Magnetization measurements were performed using a Quantum Design SQUID magnetometer over a temperature range of 2-800\,K using an applied magnetic field of 1000\,Oe. This was done in two separate measurements for low-temperature (2-340\,K) and high-temperature (300-800\,K) regimes. In the latter case, the powder sample was sealed inside an evacuated quartz tube.

\begin{figure}
\includegraphics[width=0.9\columnwidth]{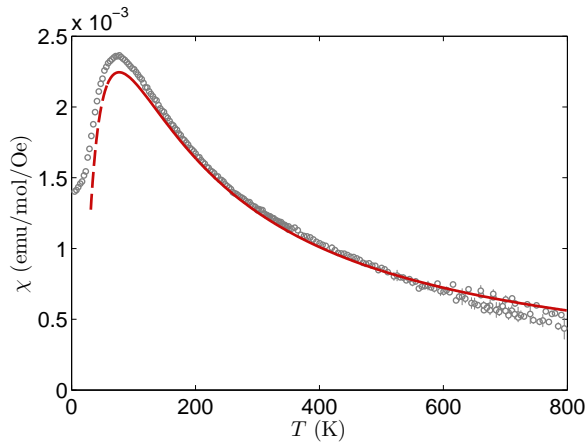}
\caption{(Color online)
(a) Magnetization measurements of \SCTO\ in a 1000\,Oe applied field. The red line shows HTE calculations for $J_1 = 7.18$\,meV, $J_2 = 0.025$\,meV, and $g$-factor of 2.3. Dashed line indicates a 10\% difference between the Pade approximation and 8th order series expansion.
}
\label{figS1}
\end{figure}

In Fig.~\ref{figS1} we show our magnetization measurements as a function of temperature performed on a \SCTO\ sample. We observe a broad peak around 74\,K which is characteristic of low-dimensional Heisenberg antiferromagnets and is consistent with previous reports \cite{iwanaga-jssc-1999,koga-jpn-2014}. Previous estimates of the exchange parameters were based on modelling magnetic susceptibility measurements using quantum Monte Carlo simulations with $J_1 = 6.4$\,meV \cite{koga-jpn-2014}.

Using thermodynamic perturbation expansions for the $J_1$-$J_2$ system, it is possible to calculate the high-temperature expansion (HTE) of the magnetic susceptibility. Using the exchange parameters obtained from fitting inelastic neutron measurements by spin-wave theory with a correction for quantum effects using series-expansion, we have performed such a calculation using algorithms developed in Ref.~\cite{schmidt-prb-2011}. Figure~\ref{figS1} shows that the HTE susceptibility obtained using [4,4] Pad\'{e} approximant describes our data very well across a wide range of temperatures. Furthermore, we can estimate the Weiss temperature in the mean-field approximation from $\theta = -S(S+1)/(3 k_{\rm B})\sum z_i J_i$ to be $-$83.6(6)\,K, which agrees reasonably well with the independently reported value of $-$97\,K for \SCTO\ \cite{iwanaga-jssc-1999}.

\section{Temperature dependence of excitations}

\begin{figure*}[htb]
\includegraphics[bb= 50 50 580 230,clip=,width=0.9\textwidth]{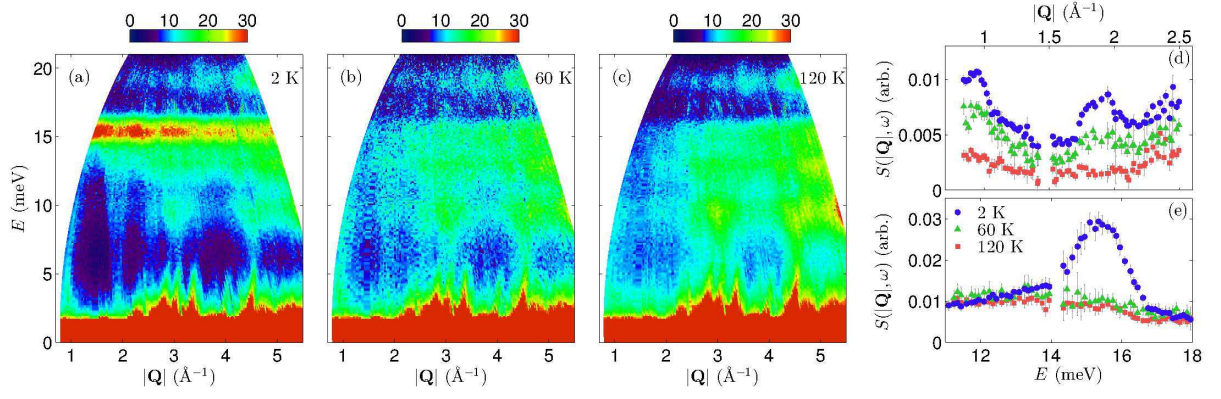}
\caption{(Color online)
Inelastic powder spectra of $S(\Qb,\omega)$ maps collected at 2, 60 and 120\,K (a--c). Representative cuts at constant energy and wavevector are shown in panels (d) $E=5$\,meV and (e) $\Qb=1.7$\,\AA$^{-1}$, respectively. In panel (d), we have displaced the data vertically for clearer presentation.
}
\label{figS2}
\end{figure*}

In the main article we have shown our analysis of the $\chi''(\Qb,\omega)$ which has been calculated through a subtraction of 120\,K measurements from 2\,K data. In Fig.~\ref{figS2} we instead show the $S(\Qb,\omega)$ measurements at 2, 60 and 120\,K. The base temperature measurements in Fig.~\ref{figS2}(a) show that scattering from the phonons is noticeable at larger \Qb. A flat band is observed around 19\,meV which we ascribe to be originating from the lattice as its intensity increases with \Qb. At 60\,K, see Fig.~\ref{figS2}(b), we observe that the intensity of the strong magnetic scattering band at 15.4\,meV is greatly suppressed. However, we still find steeply rising excitations around 1 and 1.8\,meV emanating from the magnetic Bragg positions. In Fig.~\ref{figS1}(d) we plot cuts through this dispersion at 5\,meV.
This temperature corresponds to approximately 2\TN. In Fig.~\ref{figS2}(c), we observe that phonon scattering has increased and is more prominent at larger \Qb. The magnetic fluctuations are largely gone (smeared out), which is also evident from the \Qb- and energy-cuts in Figs.~\ref{figS2}(d) and \ref{figS2}(e). The observation of magnetic fluctuations well above \TN\ which is characteristic of low-dimensional magnets, such as for example CFTD \cite{ronnow-prl-2001}. With increasing temperature, we still have magnetic scattering but these become increasingly uncorrelated and by the sum-rule are smeared out over all reciprocal space.

\end{document}